\documentstyle [12pt]{article}

\begin{document}

\vspace{2mm}

\begin{flushright}
Preprint MRI-PHY/5/95 \\

hep-th/9505167, May 1995 
\end{flushright}

\vspace{2ex}

\begin{center}
{\large \bf Short Distance Repulsive Gravity \\ 

\vspace{2ex} 

            as a Consequence of Non Trivial \\
 
\vspace{2ex} 

            PPN Parameters $\beta$ and $\gamma$ } \\

\vspace{6mm}
{\large  S. Kalyana Rama and Sasanka Ghosh}
\vspace{3mm}

Mehta Research Institute, 10 Kasturba Gandhi Marg, 

Allahabad 211 002, India. 

\vspace{1ex}
email: krama@mri.ernet.in , sasanka@mri.ernet.in  \\
\end{center}

\vspace{4mm}

\begin{quote}
ABSTRACT.  

We look for a graviton-dilaton theory which can predict non trivial 
values of the PPN parameters $\beta$ and/or $\gamma$ for a charge 
neutral point star, without any naked singularity. With the potential for 
dilaton $\phi$ set to zero, it contains one arbitrary function 
$\psi(\phi)$. Our requirements impose certain constraints on $\psi$, 
which lead to the following generic and model independent novel results: 
For a charge neutral point star, the gravitational force becomes 
repulsive at distances of the order of, but greater than, 
the Schwarzschild radius $r_0$. There is also no horizon for $r > r_0$. 
These results suggest that black holes are 
unlikely to form in a stellar collapse in this theory.  

\end{quote}

\newpage

\vspace{2ex} 
 
In Einstein's theory of general relativity, the gravitational field of 
a charge neutral point star is described by static spherically 
symmetric Schwarzschild solution. Two of the parametrised post 
Newtonian (PPN) parameters $\beta$ and $\gamma$ 
can be calculated from such a solution. The parameter $\beta$ is 
a measure of non linearity in the superposition law for gravity, and 
$\gamma$ is a measure of the space time curvature \cite{will}. In 
Einstein's theory $\beta = \gamma = 1$. Experimentally, they can be 
measured by perihelion shifts of planets and Shapiro time delay, and 
are given by $\frac{1}{3} (2 + 2 \gamma - \beta) = 1.003 \pm .005$ 
and $\gamma = 1 \pm .001$, in good agreement with Einstein's theory.  

However, for various reasons as described in detail in \cite{will}, it 
is worthwhile to consider alternative theories of gravity, the popular  
ones being the Brans-Dicke (BD) theory with a constant parameter 
$\omega_{bd} > 0$ \cite{bd} 
and the string theory. A common feature among these 
theories is the presence of a scalar field $\phi$, often called dilaton.  
The acceptable static spherically symmetric solutions for 
a charge neutral point star in BD and string theory, including only 
the graviton and the dilaton field, are all found to give 
$\beta = \gamma = 1$. There are more general 
solutions \cite{buchdahl} giving $\beta = 1$ and 
$\gamma = 1 + b$, but they always have naked curvature singularities 
proportional to $b^2$ \cite{kppn} and, hence, are unacceptable. 
Therefore these theories cannot predict non trivial values for $\beta$ 
and/or $\gamma$ for a charge neutral point star without introducing 
naked singularities.  

In this letter, we look for a graviton-dilaton theory which can predict  
non trivial values for $\beta$ and/or $\gamma$ for a charge neutral 
point star without introducing naked singularities. The most general 
graviton-dilaton theory is the generalised BD theory which, with 
the dilaton potential set to zero, contains one arbitrary function 
$\psi(\phi)$ \cite{gbd,will}. We study static 
spherically symmetric solutions in this theory and require that the 
PPN parameters $\beta$ and/or $\gamma$ be non trivial for a charge 
neutral point star, and that there be no naked singularities. 
See footnote 6 below. 
These requirements impose certain constraints on $\psi$ 
which, to the best of our knowledge, are not all satisfied in 
any of the existing graviton-dilaton theories, including the recent 
string theoretic model of \cite{damour}. 

These constraints on $\psi$ lead to generic and model independent novel 
results. For a charge neutral point star, the gravitational force 
becomes repulsive at distances of the order of, but greater than, 
the Schwarzschild radius $r_0$. As a consequence, test particles 
with non zero rest mass cannot reach $r_0$. By construction, there 
is no naked singularity, but now there is also no horizon for $r > r_0$. 
These results, as will be discussed later, 
suggest that black holes are unlikely to form 
in a stellar collapse in this theory. Such non formation of black holes, 
if established rigorously, will obviate the vexing problems of black 
hole singularities and information loss due to Hawking radiation. 
We do not understand at present the photon propagation for 
$r \le r_0$ and, hence, are unable 
to determine the space time structure completely. 
But, this may not be an issue if physical stars do not collapse 
to point objects. 
Our theory also has novel, model independent cosmological features 
which are presented in a companion letter \cite{kbbts}. 

It is interesting to note that partially similar phenomena also appear 
in two entirely different models \cite{satya,moffat}. 
In \cite{satya}, a Higgs potential for a complex BD type scalar results 
in a repulsive gravity. In the `non symmetric gravitational theory' of 
\cite{moffat} the gravitational force vanishes at $r = 0$ and 
the horizon is absent, leading its authors to conclude that there are 
no black holes. But this conclusion is debatable \cite{ori}. Also, this 
theory predicts that space is anisotropic and birefringent \cite{biref}. 

Consider now the most general action for the graviton and dilaton, 
including the matter which determine the space time structure, and 
the world line actions for the test particles \cite{bd,dicke,gbd} - in 
our notation, the signature of the metric is $(- + + +)$ and 
$\bar{R}_{\mu \nu \lambda \tau} = \frac{1}{2} \frac{\partial^2 
\bar{g}_{\mu \lambda}} {\partial x^{\nu} \partial x^{\tau}} + \cdots$ :  
\begin{eqnarray}
S & = & \int d^4 x \sqrt{- \bar{g}} e^{\bar{\psi}} \left( \bar{R} 
- \bar{\omega} (\bar{\nabla} \bar{\phi})^2  + \bar{V} \right)  
+ S_M (\{\chi\}, \{\bar{\sigma}\}, \bar{g}_{\mu \nu}) \nonumber \\
& & + \sum_i m_i \int \left( - e^{\bar{\sigma}_i} \bar{g}_{\mu \nu} 
d x^{\mu}_i d x^{\nu}_i \right)^{\frac{1}{2}} \; , \label{gen} 
\end{eqnarray}
where $\bar{\psi}, \; \bar{\omega}, \; \bar{V}$, and $\bar{\sigma}_i$  
are functions of the scalar $\bar{\phi}$, related to the dilaton (see 
below), $S_M$ is the matter action with $\{\chi\}$ and 
$\{\bar{\sigma}\}$ 
denoting symbolically matter and their dilatonic couplings, 
$m_i$'s are constants, and  the sum is over different types of 
test particles at least some of which have, by assumption, non zero rest 
mass, {\em i.e.} $m_i \ne 0$ at least for some $i$.\footnote{In 
the following, unless otherwise stated, test particles refer to those 
only with non zero rest mass.} 
We also assume that matter constituents have non zero rest mass. 
Due to the dilatonic couplings, the test particles 
feel dilatonic forces and, hence, will not fall 
freely along the geodesics of $\bar{g}_{\mu \nu}$ \cite{dicke,gbd}. 

In the following, we will set the dilaton potential 
$\bar{V} = 0$.\footnote{Static spherically symmetric graviton-dilaton 
solutions, with $\bar{\psi} = \bar{\phi}$ and $\bar{\omega} = 1$, have 
been studied in \cite{kppn,hh} for different choices 
of $\bar{V}$.} When $\{\bar{\sigma}\}$ and $\bar{\sigma}_i$'s 
are different, it amounts to 
introducing a composition dependent `fifth force' (see 
\cite{dicke,gbd,will}). Hence, we take 
$\{\bar{\sigma}\} = \bar{\sigma}_i = \bar{\sigma}$ 
for all $i$, thus coupling dilaton universally to all matter and 
test particles. 

Now, the action in (\ref{gen}) appears to have three arbitrary functions 
$\bar{\psi}, \; \bar{\omega}$ and $\bar{\sigma}$. However, first by 
a $\bar{\phi}$-dependent conformal transformation of $\bar{g}_{\mu \nu}$, 
and then by a scalar redefinition of $\bar{\phi}$, 
two of these arbitrary functions can be gotten 
rid of. The action (\ref{gen}) can then be written as: 
\begin{eqnarray}
S & = & \int d^4 x \sqrt{- g} 
\left( R + \frac{1}{2} (\nabla \phi)^2 \right) 
+ S_M (\{\chi\}, e^{- \psi} g_{\mu \nu})  \nonumber \\
& & + \sum_i m_i \int \left( - e^{- \psi} g_{\mu \nu} 
d x^{\mu}_i d x^{\nu}_i \right)^{\frac{1}{2}} \; , \label{sein}
\end{eqnarray}
where the redefined scalar is referred to as the dilaton $\phi$ and 
the arbitrary function $\psi (\phi)$ charcterises our 
graviton-dilaton theory. $\psi$ cannot be gotten rid of by any further 
field redefinitions (except when {\em all} matter constituents and 
test particles have zero 
rest mass which, by assumption, is not the case here).  
Note that in (\ref{sein}) the curvature scalar $R$ 
appears canonically. For this reason, $g_{\mu \nu}$ is often referred 
to as `Einstein metric'. However, matter and 
test particles couple to dilaton 
now and feel both the gravitational and the dilatonic forces. Hence, 
they do not fall freely along the geodesics of $g_{\mu \nu}$. 

The action in (\ref{sein}) can be written equivalently in terms of 
the metric 
\begin{equation}\label{conf}
\hat{g}_{\mu \nu} = e^{- \psi} g_{\mu \nu} \; .
\end{equation}
It then becomes 
\begin{eqnarray}
S & = & \int d^4 x \sqrt{- \hat{g}} e^{\psi} 
\left( \hat{R} - \omega (\hat{\nabla} \phi)^2 \right) 
+ S_M (\{\chi\}, \hat{g}_{\mu \nu})  \nonumber \\
& & + \sum_i m_i \int \left(- \hat{g}_{\mu \nu} 
d x^{\mu}_i d x^{\nu}_i \right)^{\frac{1}{2}} \; , \label{spsi} 
\end{eqnarray}
where $\omega = \frac{1}{2} (3 \psi_{(1)}^2 - 1)$. Here 
$\psi_{(n)} \equiv \frac{d^n \psi}{d \phi^n}$, the $n^{th}$ derivative 
of $\psi$ with respect to $\phi$. In (\ref{spsi}) and in the following, 
hats denote quantities involving $\hat{g}_{\mu \nu}$. 
Note that in (\ref{spsi}) the curvature scalar $\hat{R}$ 
does not appear canonically. However, matter and test particles 
now couple to the metric only canonically and, 
hence, fall freely along the geodesics 
of $\hat{g}_{\mu \nu}$. For this reason, we refer to $\hat{g}_{\mu \nu}$ 
as physical metric: since matter and 
test particles follow its geodesics, 
the PPN parameters or the singularities related to $\hat{g}_{\mu \nu}$ 
are the physically relevant quantities. This is the original Dicke's 
framework \cite{bd,dicke} (perhaps it is more appropriate to refer to 
$\hat{g}_{\mu \nu}$ as Dicke metric). 

It should be noted, however, that both forms of action given in 
(\ref{sein}) and (\ref{spsi}) are equivalent. It is clear that, with 
$\hat{g}_{\mu \nu}$, the physical quantities are directly obtained from 
the metric whereas, with $g_{\mu \nu}$, 
the dilatonic force must also be taken 
into account. On the other hand, it turns out that 
equations of motion are easier to solve using Einstein metric. 
In fact, in the following we will utilise both of these aspects. 

Thus, our theory is specified by the action for graviton, 
dilaton, matter, and test particles given in (\ref{sein}) 
or, equivalently, in (\ref{spsi}).  
It is characterised by one arbitrary function 
$\psi (\phi)$. Note that setting 
$\psi = \phi \; (3 + 2 \omega_{bd})^{- \frac{1}{2}}$ in (\ref{spsi}) 
one gets Brans-Dicke theory; and, setting 
$\psi = \phi$ in (\ref{spsi}) one gets the graviton-dilaton part of 
the low energy string theory\footnote{modulo the choice of 
dilatonic couplings. In the string theory literature, the action 
in (\ref{sein}), but with $\psi = 0$, and that in (\ref{spsi}), 
but with $\psi = \phi$, have both been used often.}. 

The equations of motion obtained from (\ref{sein}) are 
\begin{eqnarray}
2 R_{\mu \nu} & = & - \nabla_{\mu} \phi \nabla_{\nu} \phi 
- e^{- \psi} \left( T_{\mu \nu} 
- \frac{g_{\mu \nu}}{2} e^{- \psi} T \right) \nonumber \\
2 \nabla^2 \phi & = & \psi_{(1)} e^{- 2 \psi} T \; , 
\label{stareom}
\end{eqnarray}
where \[
T_{\mu \nu} \equiv \frac{2}{\sqrt{- \hat{g}}}  
\frac{\delta S_M}{\delta \hat{g}^{\mu \nu}} \; \; \; \; 
{\rm and} \; \; \; \; T \equiv \hat{g}_{\mu \nu} T^{\mu \nu} 
\]
are physical energy-momentum tensor of the matter and its 
trace respectively. 

We now consider static spherically symmetric vacuum solutions. 
Hence, we set $T_{\mu \nu} = T = 0$ in (\ref{stareom}), and 
first solve for $\phi$ and $g_{\mu \nu}$. We then obtain 
the physical metric $\hat{g}_{\mu \nu}$ using (\ref{conf}) 
and the physical curvature scalar $\hat{R}$ using 
\begin{equation}\label{rhat}
\hat{R} = e^{\psi} (R + \frac{3}{2} (\nabla \psi)^2 
- 3 \nabla^2 \psi) \; , 
\end{equation}
and study the physical PPN parameters and the singularities.  

With $\phi = \phi(r)$, the equations of motion in the gauge 
$d s^2 = - f(r) d t^2 + f^{- 1} (r) d r^2 + h^2 (r) d \Omega^2$, 
where $d \Omega^2$ is the line element on an unit sphere, become 
\begin{eqnarray}\label{eqn}
\frac{ (f h^2)''}{2} - 1 = (f' h^2)' & = & (\phi' f h^2)' = 0 
\nonumber \\ 
4 h'' + h \phi'^2 & = & 0 
\end{eqnarray}
where $'$ denotes $r$-derivatives. 
The most general solution to (\ref{eqn}) is \cite{buchdahl}: 
\begin{equation}\label{soln}
f = Z^a \; , \; \; h^2 = r^2 Z^{1 - a} \; , \; \; 
e^{\phi - \phi_0} = Z^b 
\end{equation}
where $Z \equiv ( 1 - \frac{r_0}{r} )$ and $\phi_0, \; r_0, \; a$, and 
$b$ are constants with $a^2 + b^2 = 1$. The constant $r_0$ is 
the Schwarzschild radius and is 
proportional to the physical mass $M$ of the point star and, 
hence, is positive. The constant $b$ can be positive or negative, but 
we take the constant $a$ to be positive so that, if $b = 0$ then $a = 1$ 
and, with $r_0 = 2 M$, one gets the standard Schwarzschild solution. 
Note that $Z$ remains positive and non vanishing for $r > r_0$. 

The physical metric $\hat{g}_{\mu \nu}$, using (\ref{conf}), is given by  
\begin{equation}\label{metric}
d \hat{s}^2 = - e^{- \psi} f d t^2 + \frac{e^{- \psi}}{f} d r^2 
+ e^{- \psi} h^2 d \Omega^2 \; . 
\end{equation}
Rewriting it in the form $d \hat{s}^2 = - \hat{f}(\rho) d t^2 
+ \hat{F}(\rho) (d \rho^2 + \rho^2 d \Omega^2)$, where \\
$r = \rho ( 1 + \frac{r_0}{4 \rho} )^2$, 
and expanding $\hat{f}$ and $\hat{F}$ in the limit 
$\rho \to \infty$, one obtains the physical mass $M$ and the 
PPN parameters $\beta$ and $\gamma$, defined as follows \cite{will}: 
\[
\hat{f} = 1 - \frac{2 M}{\rho} + \frac{2 \beta M^2}{\rho^2} 
+ \cdots \; , \; \; 
\hat{F} =  1 + \frac{2 \gamma M}{\rho} + \cdots \; .
\] 
After a straightforward algebra we find, using (\ref{soln}), that 
\begin{eqnarray}\label{ppnpsi}
2 M & = & (a - b \psi_{(1)} (\phi_0) ) \; r_0 \nonumber \\
\beta & = & 
1 - \frac{b^2 r_0^2} {4 M^2} \psi_{(2)} (\phi_0) \nonumber \\
\gamma & = & 1 + \frac{b r_0}{M} \psi_{(1)} (\phi_0) \; , 
\end{eqnarray}
where $\phi_0 = \phi |_{r \to \infty}$, and that the physical curvature 
scalar 
\begin{equation}\label{rhat1}
\hat{R} = \frac{b^2 r_0^2}{2 r^4} (3 \psi_{(1)}^2 - 6 \psi_{(2)} - 1) 
e^{\psi} Z^{a - 2} \; . 
\end{equation}

As mentioned above, $b = 0$ gives the standard Schwarzschild solution. 
Taking $\psi = \phi$, which corresponds to the low energy string 
theory, we get 
\[
2 M = (a - b) r_0 \; , \; \; 
\beta = 1 \; , \; \; \gamma = 1 + \frac{b r_0}{M} \; . 
\]
Let $b \ne 0$ so that $\gamma$ is non trivial. 
Then, $a - b > 0$ since the physical mass $M$ must be 
positive. Therefore, the $tt$-component $\hat{g}_{tt} = - Z^{a - b}$ 
vanishes at $r_0$. However, the physical curvature scalar 
$\hat{R} = b^2 r_0^2 r^{- 4} Z^{a + b - 2}$ diverges at 
$r_0$ since $a + b - 2 < 2 (a - 1) < 0$.\footnote{Here we have used 
$a - b > 0$ and, since $b \ne 0, \; a = \sqrt{1 - b^2} < 1 $.} 
This singularity will be present for 
a charge neutral point star if $b \ne 0$; or equivalently, if the PPN 
parameter $\gamma \ne 1$. Also, as shown in \cite{kppn}, this 
singularity is naked and, hence, physically unacceptable. Similar 
situation arises in Brans-Dicke theory also. Thus these theories cannot 
predict non trivial $\beta$ or $\gamma$ for a charge neutral point star, 
without implying a naked singularity. 

We would like to see if it is possible to find a graviton-dilaton 
theory which does not have this problem. That is, to find a function 
$\psi$ such that one gets a non trivial value for $\beta$ and/or 
$\gamma$ for a charge neutral point star, without any naked 
singularities. We take $b \ne 0$ and, hence, $a = \sqrt{1 - b^2} < 1 $ 
from now on. The physical mass $M$ will be positive, and $\beta$ and/or 
$\gamma$ will be non trivial and lie within the experimental bounds 
if $\psi$ satisfies 
\begin{equation}\label{c1}
a - b \psi_{(1)} (\phi_0) > 0 \; , \; \; 
b^2 \psi_{(2)} (\phi_0) = \delta \; , \; \; 
b \psi_{(1)} (\phi_0) = \epsilon \; , 
\end{equation}
where $\delta$ and $\epsilon$ are $< 10^{- 3}$, and at least 
one of them is nonzero. 

Consider the singularities. For them to be absent, not only $\hat{R}$ 
but {\em all} other curvature invariants must also be finite. As shown 
in the appendix, any curvature invariant is of the form given in 
(\ref{app1}). Now, for $r_0 \le r \le \infty, \; \phi$ ranges from 
$\phi_0$ to $\pm \infty$ depending on the sign of $b$. However, we would 
like the singularities to be absent in the region $r \ge r_0$, for 
either sign of $b$. Hence, we take $- \infty \le \phi \le \infty$. 
Then, from (\ref{app1}) follows a necessary condition for the absence 
of singularities for $r \ge r_0$:
\begin{equation}\label{c2} 
\psi_{(n)} (\phi) \equiv \frac{d^n \psi}{d \phi^n} 
= (finite) \; \; \; \; \forall \; n \ge 1 
\; \; {\rm and} \; \; - \infty \le \phi \le \infty \; .
\end{equation}
With equation (\ref{c2}) is satisfied, the necessary and sufficient 
condition for the absence of singularities at $r = r_0$ is that 
\begin{equation}\label{psilim}
\lim_{r \to r_0} e^{\psi} Z^{a - 2} = \lim_{\phi \to \pm \infty} 
e^{\psi - \frac{a - 2}{|b|} \; |\phi|} = (finite) 
\end{equation}
where, using (\ref{soln}), $Z$ is written as a function of $\phi$ only,  
valid for either sign of $b$. Equations (\ref{c2}) and (\ref{psilim}) 
then uniquely imply that 
\[
\lim_{\phi \to \pm \infty} \psi = - \frac{l}{|b|} \; |\phi| 
\; \; , \; \; \; l \ge 2 - a \; , 
\]
which can be written equivalently as 
\begin{equation}\label{c3}
\lim_{\phi \to \pm \infty} \psi = - \lambda \; |\phi| 
\; \; , \; \; \; {\rm where} \; \; 
\lambda \ge \frac{2 - a}{\sqrt{1 - a^2}} \ge \sqrt{3} 
\end{equation} 
because $0 < a < 1$. 
From equations (\ref{c2}) and (\ref{c3}) it now follows that $\psi$ 
has a finite upperbound, {\em i.e.} $\psi \le \psi_{max} < \infty$. 
Hence, $e^{\psi}$ is finite for $r > r_0$. Since $Z$ is non vanishing 
for $r > r_0$, it follows from (\ref{c2}) and (\ref{app1}) 
that the singularities are absent for $r > r_0$ as well. 

Therefore, the requirements on $\psi$ are that it satisfy equations 
(\ref{c1}), (\ref{c2}), and (\ref{c3}). The corresponding 
graviton-dilaton theory can predict non trivial values of $\beta$ 
and/or $\gamma$ for a charge neutral point star without introducing any 
singularities for $r \ge r_0$. An example of such a function $\psi$ is 
given by $\psi = - \lambda \sqrt{(\phi - \phi_1)^2 + c^2}$ where 
$\phi_1$ and $c^2$ are constants, $\lambda \ge \sqrt{3}$ and the square 
root is to be taken with positive sign. This function, for a suitable 
choice of the constants $\phi_1$ and $c$, obviously satisfies equations 
(\ref{c1}), (\ref{c2}) and (\ref{c3}). Many other examples are easy 
to obtain. 

For any function $\psi$ satisfying only equations (\ref{c1}), 
(\ref{c2}), and (\ref{c3}), consider now the physical metric 
$\hat{g}_{\mu \nu}$, in particular its $tt$-component 
$\hat{g}_{tt} = - e^{- \psi} Z^a$, \\ $0 < a < 1$. As $r$ decreases 
from $\infty$ to $r_0$, $( - \hat{g}_{tt} )$ decreases from $1$, reaches 
a minimum at $r_{min} > r_0$, and then diverges to infinity at $r_0$, 
always remaining positive and non vanishing for $r_0 \le r \le \infty$. 
This follows due to the following four reasons. 
(1) Since the physical mass $M$ is positive, 
\begin{equation}\label{gttinf}
- \hat{g}_{tt} = 1 - \frac{2 M}{r} + {\cal O}(\frac{1}{r^2}) \; < 1  
\; \; {\rm as} \; \; r \to \infty \; ; 
\end{equation}
(2) $e^{- \psi}$, which is non negative, never vanishes because 
$\psi \le \psi_{max} < \infty$; (3) $Z > 0$ for $r > r_0$; and (4) 
\begin{equation}\label{gtt0}
\lim_{r \to r_0} ( - \hat{g}_{tt} ) = Z^{a - l} 
\ge \left( 1 - \frac{r_0}{r} \right)^{2 (a - 1)} \to \infty 
\end{equation}  
because $l \ge 2 - a$, and $a < 1$. 
The exact value of $r_{min}$ is determined from 
$| b \psi_{(1)}(r_{min}) | = a$. It depends on the details 
of $\psi$, but is always greater than, and typically of the order of, 
the Schwarzschild radius $r_0$. 

Qualitatively speaking, the slope of the curve $( - \hat{g}_{tt} )$ vs. 
$r$ indicates the nature of the gravitational force. Thus, for 
the standard Schwarzschild black hole where 
$( - \hat{g}_{tt} ) = 1 - \frac{2 M}{r}$, 
the force is always attractive. In our case, then, 
the $r$-dependence of $( - \hat{g}_{tt} )$ described above 
indicates that the gravitational force is attractive for $r > r_{min}$, 
vanishes at $r_{min}$, and becomes repulsive for $r < r_{min}$. 

The repulsive force can also be seen by studying the geodesic motion of 
a radially incoming test particle with non zero rest mass. For a metric 
given by $d s^2 = - g_0 d t^2 + g_1 d r^2 + g_2 d \Omega^2$, where 
$g_0, \; g_1$, and $g_2$ are functions of $r$ only, the radial geodesic 
equation becomes 
\begin{equation}\label{acc}
r_{pp} + \frac{g'_1 r_p^2}{2 g_1} + \frac{g'_0}{2 g_1 g_0^2} = 0 
\; , \; \; t_p = \frac{1}{g_0} \; , 
\end{equation}
where $(\;)' \equiv \frac{d (\;)}{d r}$ and 
$(\;)_p \equiv \frac{d (\;)}{d p}$. 
Equation (\ref{acc}) can be integrated twice to get 
\begin{equation}\label{tr}
\int dt = \int dr \sqrt{\frac{g_1}{g_0 (1 + E g_0)}} \; 
\end{equation}
where $E = - 1 + v^2$, corresponding to releasing the test particle  
at $r = \infty$ with an inward velocity $v$ (in units where velocity 
of light $= 1$). Since the test particle has non zero 
rest mass, its velocity $v < 1$ and, hence, $E < 0$. 
In our case $g_0 = - \hat{g}_{tt}$ which, as shown above, 
diverges to $\infty$ at $r_0$. Therefore, the denominator in (\ref{tr}) 
vanishes at some $r_t$, where $1 + E g_0(r_t) = 0$ and 
$r_0 < r_t < r_{min}$, indicating that $r_t$ is the turning point. 
Analysis of equation (\ref{acc}) then shows that the test particle 
travels outwards after reaching $r_t$. It is clear that such a turning 
point exists irrespective of the value of $v \; ( < 1)$ or, equivalently, 
the initial energy of the test particle.  This shows that test particles 
feel a repulsive gravitational force near $r_0$.\footnote{Contrast this 
with the Schwarzschild black hole where $g_0 = 1 - \frac{r_0}{r} \le 1$: 
the factor $1 + E g_0$ never vanishes and, hence, there is no turning 
point.} In Einstein frame, where the action is given by (\ref{sein}), 
this repulsion can be thought of as arising due to dilatonic force. 

The repulsive gravitational force has novel implications for 
the collapse of a star. Consider a non rotational collapse and 
assume, in the following, that the star is so massive that its 
matter pressure is insufficient to prevent gravitational collapse. 

In Einstein's theory, such a star collapses completely, becoming  
a black hole \cite{hawking1} (see also \cite{chris}).  
In \cite{hawking2}, it has been shown 
that if a black hole forms in a stellar collapse in 
Brans-Dicke theory, then the dilaton will become constant everywhere. 
Thus, the original dilaton field, if any, of a star will all be 
radiated away when it collapses and becomes a black hole. But, 
reference \cite{hawking2} does not address the question of whether 
or not a star in Brans-Dicke theory {\em will} collapse to become 
a black hole. Recently, this question has been answered affirmatively 
in \cite{shapiro}, where the authors show that the star 
in Brans-Dicke theory indeed collapses and becomes a black hole, 
radiating away all the dilaton field in the process.\footnote{Thus, 
a charge neutral star can have non trivial $\beta$ and $\gamma$. 
When it collapses and forms a black hole, the dilaton evaporates 
away producing $\beta = \gamma = 1$. This, then, weakens 
the original motivation for the construction of the present 
graviton-dilaton theory. Nevertheless, we wish to continue its 
study in view of the simplicity of the initial requirements 
and the consequent generic novel features.} 
This is the standard collapse scenario. 

The graviton-dilaton theory described in this paper suggests 
naturally the following collapse scenario. Let $r_*$ be the radius 
of the star. The solutions presented here may describe 
the gravitational field outside $r_*$, while the fields inside may 
be obtained by solving the matter coupled equations and imposing 
appropriate boundary conditions at $r_*$. At equilibrium, $r_*$ 
cannot be less than $r_{min}$ since the gravitational force becomes 
repulsive for $r < r_{min}$, causing the star to expand. During 
a collapse, when $r_*$ becomes $< r_{min}$, the gravitational 
repulsion may halt and reverse the collapse, eventually stabilising 
$r_*$ around $r_{min}$. Since there is no horizon for $r > r_{min} \; 
( > r_0)$, a stellar collapse in this theory may not result in 
the formation of a black hole, even if the star is supermassive. 
Note that the repulsive gravitational force is crucial in this 
scenario. 

Is this scenario plausible? We think that the answer is yes, for 
the following reasons. The standard scenarios in 
\cite{hawking1}-\cite{hawking2} do not apply to our case. 
In Einstein frame, the gravitational repulsion can be thought 
of as arising due to the dilatonic coupling of matter. The matter 
in \cite{hawking1,chris} has no dilatonic coupling and, hence, no 
repulsive force will ever arise, making the analyses of 
\cite{hawking1,chris} inapplicable to our case. 
An important reason for 
the numerical simulation of \cite{shapiro} itself is that 
reference \cite{hawking2} sidesteps the question of whether 
or not a collapse in Brans-Dicke theory results in a black hole. 
The same objection and the fact that gravity remains attractive 
in BD theory make the analysis of \cite{hawking2} 
inapplicable to our case also. 

Consider now the relevant points from the analysis 
of \cite{shapiro}. The collapse simulation is carried out for  
$\omega_{bd} \ge 0$, where $\omega_{bd}$ is the constant 
BD parameter. It is carried out for $\omega_{bd} < -2$ also, 
but this will not be relevant to our purpose. The simulation 
has not been carried out for $-2 < \omega_{bd} < 0$ due to 
the subtleties arising from the singular behaviour of the factor 
$(2 \omega_{bd} + 3)^{- 1}$ appearing in the source term in 
the dilaton equation. Also, the gravitational force 
always remains attractive in BD theory.

In the present theory, the gravitational force becomes 
repulsive near $r_0$. The analog of the BD parameter 
now is 
$\omega_{bd} (\phi) = - \frac{3}{2} + \frac{1}{2 \psi_{(1)}^2}$, 
a function of dilaton \cite{kbbts}. 
Near $r_0$, where the gravitational force 
becomes repulsive, 
$- \frac{3}{2} \le \omega_{bd} (\phi) \le - \frac{4}{3}$ because 
$\psi_{(1)}^2 = \lambda^2 \ge 3$ (see equation (\ref{c3})). 
These values of $\omega_{bd} (\phi)$ 
are the ones likely to be relevant to 
our collapse scenario, and are precisely those that were not 
investigated in \cite{shapiro}. Hence, this work does not rule 
out the collapse scenario in the present theory, 
described above \cite{private}. 

Consider equations (\ref{stareom}). They describe the dynamics 
of collapse. In the dilaton equation, qualitatively speaking, 
the metric coupling in $\nabla^2 \phi$ causes the dilaton to 
evaporate away during the collapse to a black hole, whereas 
the matter/self-coupling term $\psi_{(1)} e^{- 2 \psi} T$ 
acts as a source for $\phi$. 

Consider these terms. For a metric and the dilaton of the form 
similar to that given in (\ref{soln}), and for a slow enough 
collapse, the terms in $\nabla^2 \phi$ diverge near $r_0$ at most 
as $Z^{- 2}$, where $Z = 1 - \frac{r_0}{r}$ and $r_0$ is 
the Schwarzschild radius, now likely to depend on both 
$r$ and $t$.  In the source term, $T$ is non divergent. 
$\psi_{(1)}$ is a constant and is $< 1$ in BD theory; it is 
also a constant, but $>> 1$, near $r_0$ in our theory. The factor 
$e^{- 2 \psi}$ diverges near $r_0$ as $Z^{- 2 k}$ where, in BD 
theory, $k = \frac{b}{3 + 2 \omega_{bd}} << 1$ since $b$ is 
small and $\omega_{bd}$ is large. Thus, in BD theory, 
the gravitational effects on $\phi$ can be expected to dominate 
the source effects. This can also be seen in \cite{shapiro}, 
where the dilaton initially grows in magnitude and starts 
evaporating away just when a horizon is forming (see figures 1-5  
in \cite{shapiro}). In the present theory, on the other 
hand, for any function $\psi$ that 
satisfies the constraints given in (\ref{c1}), (\ref{c2}), and 
(\ref{c3}), $k = l \ge (2 - a) > 1$ since $a < 1$. It  
implies that source effects become dominant now! This makes it 
plausible that despite the metric induced evaporation, 
the dilaton may still grow in magnitude. In that case, 
the gravitational repulsion would set in, and the collapse 
is likely to be halted as in the scenario proposed above. 

Because of these reasons, we think that the standard 
collapse scenario may not be applicable in the present 
graviton-dilaton theory, and that the scenario we have 
proposed may 
be the right one. The precise dynamics of collapse, however, 
needs to be understood in full detail, but its analytical 
study is quite difficult. It can, perhaps, be best understood 
by numerical simulations only as in \cite{shapiro}, and is 
presently under study. 

Consider now test particles with zero rest mass (photons), 
propagating in the space time given by (\ref{soln}). Strictly 
outside $r_0$, the nature of their propagation is similar to that in 
the Schwarzschild case. For example the redshift, as also the traversal 
time measured in a laboratory, of a photon travelling from any 
$r_i > r_0$ to $r_f > r_i$ is finite. Since $\hat{g}_{tt}$ is 
non vanishing and there are no singularities in this region, one 
concludes that there is no horizon for $r > r_0$. However, we do not 
understand at present the photon propagation for $r \le r_0$. 
The problem we face is that the dilaton and the metric components 
become complex when $r < r_0$. Also, the factor $e^{- \psi}$ diverges 
at $r = r_0$ and, hence, the validity of the conformal transformation 
(\ref{conf}) appears to be questionable. Consequently, we are unable  
to determine completely the space time structure. 

It may turn out to be the case that in 
physical situations one need not worry about the region  
$r \le r_0$. If the collapse scenario described above holds good, 
then the radius $r_*$ of any star $> r_0$. The present solutions are 
then applicable only outside the star, {\em i.e.}\ for $r > r_* > r_0$. 
The solutions inside the star, {\em i.e.}\ for $r < r_*$ (which includes 
the region $r \le r_0$), will be quite different from the present 
ones. Then, determining the space time structure may not be 
an issue. This is unlike in the Schwarzschild case where 
a black hole can form in physical situations, {\em i.e.} $r_*$ 
can be $< r_0$ and, hence, one does need to worry about 
the region $r \le r_0$. 

To summarise, we have constructed a graviton-dilaton theory by 
requiring that it can predict non trivial values of the PPN parameters 
$\beta$ and/or $\gamma$ for a charge neutral point star, without any 
naked singularity. It contains one function $\psi (\phi)$. Our simple 
requirements constrain $\psi$ to satisfy equations (\ref{c1}), 
(\ref{c2}), and (\ref{c3}). These constraints lead to generic and model 
independent novel results as described here. The cosmological features 
of this theory are presented in a companion letter \cite{kbbts}. 

Keeping in view the simplicity of the initial requirements, 
and the resulting novel features that, if established 
rigorously, may have far reaching consequences, we believe 
that further study of the present graviton-dilaton theory 
will be fruitful. 

\vspace{2ex} 

{\bf Acknowledgements:} 

It is a pleasure to thank H. S. Mani and T. R. Seshadri for 
numerous helpful discussions, A. Sen for comments, and 
M. A. Scheel and S. A. Teukolsky for a communication. We are 
particulary greatful to the referee for pointing out a major 
flaw in our original proof of the absence of singularities 
- correcting this flaw led to the more stringent constraint on 
$\psi$ given in (\ref{c2}), for informing us of the references  
\cite{buchdahl,chris}, and for pointing out some mistakes in 
the original version. The paper has benefited much from 
his/her comments. 

\vspace{4ex}

\begin{center}
{\bf Appendix}
\end{center}

All curvature invariants can be constructed using metric tensor, 
Riemann tensor, and covariant derivatives which contain ordinary 
derivatives and Christoffel symbols $\hat{\Gamma}_{a b}^c$. 
When the metric is diagonal, every term in 
any curvature invariant can be grouped into factors, each of which is 
of one of the following forms (no summation over repeated indices): 
(A) $\sqrt{\hat{g}^{a a} \hat{g}^{b b} \hat{g}^{c c} \hat{g}^{d d}} 
\hat{R}_{a b c d}$, (B) $\sqrt{\hat{g}^{a a} \hat{g}^{b b} 
\hat{g}_{c c}} \hat{\Gamma}_{a b}^c$, 
or (C) $\sqrt{\hat{g}^{a a}} \partial_a$. 

Taking $\hat{g}_{\mu \nu}$ given in (\ref{metric}), and the solutions 
for $f, \; h$, and $\phi$ given in (\ref{soln}), the above forms can be 
calculated explicitly. The calculation is straightforward, and 
the result is that (A), (B), and (C) can be written, symbolically, as 
\begin{eqnarray*}
& & (A) \simeq U e^{\psi} Z^{a - 2} \; , \; \; \; 
(B) \simeq V ( e^{\psi} Z^{a - 2} )^{\frac{1}{2}} \; ,  \\
& & (C)^n \cdot (A)^p (B)^q \simeq W_{n + 2} 
( e^{\psi} Z^{a - 2} )^{p + \frac{1}{2}(q + n)} \; , 
\end{eqnarray*}
where $U$ and $V$ are functions of $r, \; \psi_{(1)}$, and 
$\psi_{(2)}$ only, and $W_{k}$ are functions of $r$ and 
$\psi_{(l)}, \; \; 1 \le l \le k$. 
As a result of the way various factors 
are grouped, it turns out that, the explicit 
$r$-dependent parts in $U, \; V$, and $W_k$'s are finite for 
all $r \ge r_0$ (in fact, they diverge only at $r = 0$). 

Hence, any curvature invariant constructed from $m$ Riemann tensors, 
$n$ covariant derivatives, and the requisite number of metric tensors 
will be of the form 
\begin{equation}\label{app1}
\tilde{W} (r; \psi_{(1)}, \psi_{(2)}, \cdots, \psi_{(n + 2)}) \; 
( e^{\psi} Z^{a - 2} )^{m + \frac{n}{2}} \; 
\end{equation}
where the explicit $r$-dependent parts of $\tilde{W}$ are finite for 
all $r \ge r_0$ (in fact, they diverge only at $r = 0$). Note, as 
an example, that the curvature scalar given in (\ref{rhat1}) belongs 
to type (A), and has the above form with $m = 1$ and $n = 0$.

\end{document}